\documentclass[preprint,showpacs,preprintnumbers,amsmath,amssymb]{revtex4}
\usepackage{graphicx}
\usepackage{dcolumn}
\usepackage{bm}
\usepackage{color}
\usepackage{hhline}

\begin{document}

\title{Orbital dependent electron tunneling within the atom superposition approach: Theory and application to W(110)}

\author{Kriszti\'an Palot\'as}
\email{palotas@phy.bme.hu}
\affiliation{Budapest University of Technology and Economics,
Department of Theoretical Physics, Budafoki \'ut 8., H-1111 Budapest, Hungary}

\author{G\'abor M\'andi}
\affiliation{Budapest University of Technology and Economics,
Department of Theoretical Physics, Budafoki \'ut 8., H-1111 Budapest, Hungary}

\author{L\'aszl\'o Szunyogh}
\affiliation{Budapest University of Technology and Economics,
Department of Theoretical Physics and Condensed Matter Research Group of the Hungarian Academy of Sciences,
Budafoki \'ut 8., H-1111 Budapest, Hungary}

\date{\today}

\begin{abstract}

We introduce an orbital dependent electron tunneling model and implement it within the atom superposition approach for simulating
scanning tunneling microscopy (STM) and spectroscopy (STS). Applying our method, we analyze the
convergence and the orbital contributions to the tunneling current and the corrugation of constant current STM images
above the W(110) surface. In accordance with a previous study [Heinze et al., Phys. Rev. B 58, 16432 (1998)],
we find atomic contrast reversal depending on the bias voltage.
Additionally, we analyze this effect depending on the tip-sample distance using different tip models, and
find two qualitatively different behaviors based on the tip orbital composition.
As an explanation, we highlight the role of the real space shape of the orbitals involved in the tunneling.
STM images calculated by our model agree well with Tersoff-Hamann and Bardeen results.
The computational efficiency of our model is remarkable as the k-point samplings of the surface and tip Brillouin zones
do not affect the computation time, in contrast to the Bardeen method.

\end{abstract}


\maketitle

\section{Introduction}

The experimental use of scanning tunneling microscopy (STM) and spectroscopy (STS) has recently gained a boost. The explanation of
experimentally observed effects is not straightforward without a proper theoretical support \cite{hofer03rmp,hofer03pssci}.
One direction of recent theoretical research is focused on extracting surface local electronic properties from experimental STS
data \cite{ukraintsev96,koslowski07,passoni09,ziegler09,koslowski09}, which is the convolution of tip and sample electronic
structures. Another research direction is concerned with the simulation of STM and STS by using different models mostly based
on electronic structure data obtained from first principles.

Much work has been devoted to analyze the electron tunneling properties depending on the scanning tip.
In Ref.\ \cite{hofer05sts} a theoretical method has been presented to separate the tip and sample contributions to STS.
Ness and Gautier studied different metal tips and their interaction with metal surfaces in a tight-binding framework
\cite{ness95jpcm1,ness95jpcm2,ness95prb}. Chen and Sacks investigated the effect of the tip orbitals
on the corrugation of constant current STM images theoretically \cite{chen90,sacks00}.
While Chen pointed out that corrugation enhancement is expected for tip orbitals localized along the surface normal ($z$)
direction ($p_z$ and $d_{3z^2-r^2}$), Sacks argued that $m\ne 0$ tip states ($d_{xz}$, $d_{yz}$, $d_{xy}$, $d_{x^2-y^2}$) are
responsible for this effect. Another work of Chen explained the corrugation inversion found on low Miller index metal surfaces
by $m\ne 0$ tip states \cite{chen92}. Atomic contrast reversal has also been found above Xe atomic adsorbates \cite{mingo96}
and oxygen overlayers \cite{calleja04} on metal surfaces. It was established that the character of the contrast depends on the
tip-sample distance and on the tip geometry and electronic structure.
The quality of the tip also plays a crucial role in the inelastic tunneling spectroscopy (IETS). Studying the CO molecule on Cu
surfaces it has been found that the IETS intensity is close to the experiment including the full electronic structure of the tip
\cite{teobaldi07}, and the tip position and orbital symmetry can change the selection rules for the active vibrational modes in
IETS \cite{garcia-lekue11}.

The role of the electron orbitals has been considered in different electron transport models.
Sirvent et al.\ developed a tight-binding model based on the Keldysh formalism for calculating the conductance in atomic point
contacts and analyzed the effect of the $d$ orbitals \cite{sirvent96}.
The same methodology has been applied to STM junctions by Mingo et al.\ \cite{mingo96}.
Cerd\'a et al.\ presented an STM simulation method based on the Landauer-B\"uttiker formula \cite{buttiker85} and the
surface Green function matching technique \cite{cerda97}.
In these methods the decomposition of the conductance/current with respect to electron orbitals has been provided.

In the present work we consider a simple model for orbital dependent tunneling within the atom superposition approach for
simulating STM \cite{palotas11stm} and STS \cite{palotas12sts}. The main idea of the paper is the introduction of a
geometrical factor to account for a modified transmission due to electron orbital orientational overlap effects within a
three-dimensional (3D) Wentzel-Kramers-Brillouin (WKB)-based theory.
The reliability of this new method is demonstrated by the analysis of the tip-sample distance and
bias voltage dependent corrugation reversal effect observed on the W(110) surface, where we find excellent agreement with a
previous work \cite{heinze98}. The computational efficiency of our method, which is justified in section \ref{sec_compar}, enables
us to study this effect in much more detail. We particularly focus on tip effects, and consider ideal tip models with different
orbital symmetries and a more realistic W(110) tip. We find two qualitatively different corrugation inversion behaviors based
on the tip orbital composition. Our results indicate that anticorrugation on the W(110) surface can not only be observed at
negative bias voltages but also at positive bias at reasonably short tip-sample distances.

The paper is organized as follows: The theoretical model of the orbital dependent tunneling within the atom superposition
approach is presented in section \ref{sec_theory}.
Applying this method we investigate the convergence and the orbital contributions of the tunneling current
as well as the corrugation reversal of the W(110) surface depending on the applied bias voltage
and the tip-sample distance in section \ref{sec_res}. We also report a comparison of STM images calculated by our model to
Tersoff-Hamann \cite{tersoff83,tersoff85} and Bardeen \cite{bardeen61} results in section \ref{sec_compar}.
Summary of our findings is found in section \ref{sec_conc}.

\section{Orbital dependent tunneling model within 3D WKB theory}
\label{sec_theory}

Recently, Palot\'as et al.\ developed a 3D atom superposition approach for simulating spin-polarized STM
(SP-STM) \cite{palotas11stm} and spin-polarized STS (SP-STS) \cite{palotas12sts} based on previous theories
\cite{passoni09,wortmann01,yang02,smith04,heinze06}. In the model it is assumed that electrons tunnel through one tip apex
atom, and contributions from individual transitions between this apex atom and each of the surface atoms are summed up assuming
the one-dimensional (1D) WKB approximation for electron tunneling processes in all these transitions.
The key input is the projected electron density of states (PDOS) of the tip apex atom and of the sample surface atoms
obtained from ab initio electronic structure calculations. In the present paper first we review the tunneling current and
the differential conductance expressions based on the independent orbital approximation for the vacuum decay of electron states,
and then we extend this model to include a simple orbital dependent tunneling transmission. We consider the non-spin-polarized
part of the tunneling only, however, this theory can be applied to SP-STM and SP-STS in the future.

Assuming elastic tunneling at $T=0$ K temperature, the tunneling current at the tip position $\underline{R}_{TIP}(x,y,z)$ and at
bias voltage $V$ is given by
\begin{equation}
\label{Eq_curr}
I(x,y,z,V)=\int_{0}^{V}\frac{dI}{dU}(x,y,z,U,V)dU.
\end{equation}
The integrand is the so-called virtual differential conductance,
\begin{eqnarray}
\label{Eq_didu}
&&\frac{dI}{dU}(x,y,z,U,V)=\varepsilon^2\frac{e^2}{h}\\
&\times&\sum_{a}T(E_F^S+eU,V,d_{a}(x,y,z))n_T(E_F^T+eU-eV)n_S^{a}(E_F^S+eU).\nonumber
\end{eqnarray}
Here $e$ is the elementary charge, $h$ is the Planck constant, and $E_F^T$ and $E_F^S$ are the Fermi energies of the tip
and the sample surface, respectively. $\varepsilon^2e^2/h$ ensures that the $dI/dU$ is correctly measured in the units of $A/V$.
$\varepsilon$ has been chosen to 1 eV but its actual value has to be determined comparing simulation results to experiments
or to calculations using more sophisticated tunneling models.
Note that the exact choice of this scaling factor has absolutely no qualitative influence on the reported results, and the
comparison of current values to Bardeen results confirms our choice, see section \ref{sec_compar}.
The sum over $a$ corresponds to the atomic superposition and has to be carried out, in principle, over all surface atoms.
However, convergence tests showed that including a relatively small number of atoms in the sum provides converged $dI/dU$ values
\cite{palotas11sts}. $T(E_F^S+eU,V,d_{a}(x,y,z))$ is the energy and bias dependent tunneling transmission function, which
also depends on the distance $d_{a}(x,y,z)=\left|\underline{R}_{TIP}(x,y,z)-\underline{R}_{a}\right|$ between the
tip apex and the surface atom labeled by $a$ with position vector $\underline{R}_{a}$.
The tip and sample electronic structures are included into this model via projected DOS (PDOS) onto the atoms,
i.e., $n_T(E)$ and $n_S^{a}(E)$ denote projected charge DOS onto the tip apex and the $a$th surface atom, respectively.
They can be obtained from any suitable electronic structure method.

Taking the derivative of Eq.(\ref{Eq_curr}) with respect to the bias voltage the differential conductance is obtained.
It can be written at the tip position $\underline{R}_{TIP}(x,y,z)$ and at bias voltage $V$ as the sum of three terms,
\begin{equation}
\label{Eq_didv}
\frac{dI}{dV}(x,y,z,V)=\frac{dI}{dU}(x,y,z,V,V)+B(x,y,z,V)+D_T(x,y,z,V).
\end{equation}
Here $dI/dU$ has been defined in Eq.(\ref{Eq_didu}), and $B$ and $D_T$ are the background and tip-derivative terms,
\begin{eqnarray}
\label{Eq_B}
&&B(x,y,z,V)=\varepsilon^2\frac{e^2}{h}\\
&\times&\sum_{a}\int_{0}^{V}\frac{\partial T}{\partial V}(E_F^S+eU,V,d_{a}(x,y,z))n_T(E_F^T+eU-eV)n_S^{a}(E_F^S+eU)dU,\nonumber\\
\label{Eq_DT}
&&D_T(x,y,z,V)=-\varepsilon^2\frac{e^2}{h}\\
&\times&\sum_{a}\int_{0}^{V}T(E_F^S+eU,V,d_{a}(x,y,z))\frac{\partial n_T}{\partial U}(E_F^T+eU-eV)n_S^{a}(E_F^S+eU)dU,\nonumber
\end{eqnarray}
respectively \cite{palotas12sts}. The background term, which contains the bias-derivative of the transmission function, is usually
taken into account in recent STS theories \cite{passoni09,passoni07,donati11}, while the tip-derivative term containing the
energy derivative of the tip DOS is rarely considered in the recent literature.

In the spirit of the independent orbital approximation the transmission probability for electrons tunneling between states of
atom $a$ on the surface and the tip apex is of the simple form,
\begin{equation}
\label{Eq_Transmission}
T(E_F^S+eU,V,d_{a})=e^{-2\kappa(U,V)d_{a}}.
\end{equation}
This corresponds to a spherical exponential decay of the electron wave functions irrespective of their orbital symmetry
\cite{tersoff83,tersoff85,heinze06}.
Assuming the same effective rectangular potential barrier between the tip apex and each surface atom, the vacuum decay $\kappa$
can be written as
\begin{equation}
\label{Eq_kappa_WKB}
\kappa(U,V)=\frac{1}{\hbar}\sqrt{2m\left(\frac{\phi_S+\phi_T+eV}{2}-eU\right)},
\end{equation}
where the electron's mass is $m$, $\hbar$ is the reduced Planck constant, and $\phi_S$ and $\phi_T$ are the average electron
work function of the sample surface and the local electron work function of the tip apex, respectively. The method of determining
the electron work functions from the calculated local electrostatic potential is reported in Ref.\ \cite{palotas11stm}.

\begin{figure*}
\includegraphics[width=0.5\textwidth,angle=0]{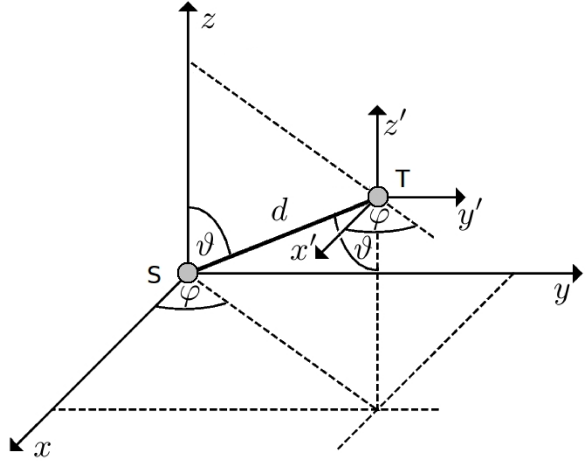}
\caption{\label{Fig1} Geometry of a general tip apex-sample atom setup.
}
\end{figure*}

Next, we extend this tunneling model by taking advantage of the orbital decomposition of the electronic structure data and
the real space shape of the electron orbitals.
The PDOS of the sample surface atoms and the tip apex can be decomposed according to orbital symmetry, i.e., real spherical
harmonics, $\beta,\gamma\in\{s,p_y,p_z,p_x,d_{xy},d_{yz},d_{3z^2-r^2},d_{xz},d_{x^2-y^2}\}$, as
\begin{eqnarray}
\label{Eq_nS_decomp}
n_S^{a}(E)=\sum_{\beta}n_{S\beta}^{a}(E),\\
n_T(E)=\sum_{\gamma}n_{T\gamma}(E).
\label{Eq_nT_decomp}
\end{eqnarray}
Similar decomposition of the Green functions has been employed in the linear combination of atomic orbitals (LCAO) scheme
by Refs.\ \cite{mingo96,sirvent96}.
Assuming such an orbital decomposition, the virtual differential conductance can be generalized as
\begin{eqnarray}
\label{Eq_didu_decomp}
&&\frac{dI}{dU}(x,y,z,U,V)=\varepsilon^2\frac{e^2}{h}\\
&\times&\sum_{a}\sum_{\beta,\gamma}T_{\beta\gamma}(E_F^S+eU,V,d_{a}(x,y,z))n_{T\gamma}(E_F^T+eU-eV)n_{S\beta}^{a}(E_F^S+eU),\nonumber
\end{eqnarray}
where, additionally to the atomic superposition (sum over $a$) we sum up each $\beta\leftrightarrow\gamma$ transition with
an orbital dependent tunneling transmission: $T_{\beta\gamma}(E,V,d_{a})$ gives the electron tunneling probability at
energy $E$ from the tip apex $\gamma$ orbital to the $\beta$ orbital of the $a$th surface atom at positive bias voltage
($V>0$), and from the $\beta$ orbital of the $a$th surface atom to the tip apex $\gamma$ orbital at negative bias ($V<0$).
$T_{\beta\gamma}$ can be defined in different ways based on physical arguments. We consider the following form,
\begin{equation}
\label{Eq_Transmission_decomp}
T_{\beta\gamma}(E_F^S+eU,V,d_{a})=e^{-2\kappa(U,V)d_{a}}t_{\beta\gamma}(\vartheta_{a},\varphi_{a})
\end{equation}
for each surface atom $\leftrightarrow$ tip apex 1D electron transition.
Here, the energy and bias dependent part corresponds to the spherical exponential decay considered in Eq.(\ref{Eq_Transmission}),
and is independent of the orbital symmetry. This is multiplied by an orbital dependent expression $t_{\beta\gamma}$,
which depends on the spatial arrangement of the sample atoms relative to the tip apex and all the orbital shapes involved in the
tunneling. The angular dependence on $\vartheta_{a}$ and $\varphi_{a}$ comes into play in the following way:
Let us consider one transition between surface atom $a$ and the tip apex along the line connecting the two atoms.
A particular geometry is shown in Figure \ref{Fig1}. For brevity, we omitted the $a$ notation of the surface atom.
For both atoms a local coordinate system can be set up, and the angular dependence of the atomic orbital wave functions
$\chi(\vartheta,\varphi)$ are defined in the corresponding coordinate system, as summarized in Table \ref{Table1}.
$\underline{R}_{TIP}-\underline{R}_{a}$ defines a vector pointing from the surface atom toward the tip apex, and it
can be represented by the $(d_{a},\vartheta_{a},\varphi_{a})$ coordinates in the spherical
coordinate system centered on the $a$th surface atom. Taking the opposite connecting vector from the tip apex toward the
surface atom, its coordinates are $(d_{a},\pi-\vartheta_{a},\pi+\varphi_{a})$ in the spherical coordinate system
centered on the tip apex.
According to Figure \ref{Fig1}, $d_{a}$, $\vartheta_{a}$, and $\varphi_{a}$ can be calculated as
\begin{eqnarray}
d_{a}=\sqrt{(x-x_{a})^2+(y-y_{a})^2+(z-z_{a})^2},\\
\vartheta_{a}=\arccos \left(\frac{z-z_{a}}{d_{a}}\right),\\
\varphi_{a}=\arccos \left(\frac{x-x_{a}}{d_{a}\sin(\vartheta_{a})}\right),
\end{eqnarray}
using the global coordinates $\underline{R}_{TIP}=(x,y,z)$ and $\underline{R}_{a}=(x_{a},y_{a},z_{a})$.
Considering above, $t_{\beta\gamma}$ accounts for the modification of the perfect spherical
exponential decay along the connecting line through the angular dependence of the atomic orbitals as
\begin{equation}
\label{Eq_Transmission_orbdep}
t_{\beta\gamma}(\vartheta_{a},\varphi_{a})=\left[\chi_{\beta}(\vartheta_{a},\varphi_{a})\right]^2\times\left[\chi_{\gamma}(\pi-\vartheta_{a},\pi+\varphi_{a})\right]^2,
\end{equation}
where $\chi_{\beta,\gamma}(\vartheta,\varphi)$ are the real spherical harmonics summarized in Table \ref{Table1}.
They were chosen in such a way that $0\le t_{\beta\gamma}\le 1$. This factor takes the effect of the directional tunneling
between real space orbitals into account. The physical motivation is the angular dependence of the electron density of the
orbitals in the given tunneling direction, which modifies the tunneling transmission. The maximum $t=1$ is obtained if the
angular distributions of the electron density according to the given orbital symmetries ($\beta,\gamma$) around both the sample
surface atom and the tip apex have maxima along the line of the tunneling direction. This is always the case for $s$-$s$ type of
tunneling irrespective of the relative position of the tip apex and sample surface atoms, i.e., we observe perfect spherical
exponential decay between tip and sample $s$ orbitals. In some particular geometries $t=1$ can be obtained even for other type of
orbitals, e.g., if the tip apex is precisely above surface atom $a$, i.e., $\vartheta_{a}=0$, then
$t_{\beta\gamma}(\vartheta_{a}=0,\varphi_{a})=1$ for all of the following combinations:
$\beta,\gamma\in\{s,p_z,d_{3z^2-r^2}\}$.
On the other hand, if the tip apex is above surface atom $a$ then orbitals with nodal planes orthogonal to
the surface have zero contribution to the tunneling from this particular surface atom, i.e., a reduced effective
tunneling transmission is obtained \cite{sacks00}.
Note that the independent orbital approximation corresponds to $t_{\beta\gamma}=1$ for all $\beta\leftrightarrow\gamma$
transitions, i.e., the same tunneling transmission is assumed between all orbitals.
Within our orbital dependent tunneling approach ideal tip models with particular orbital symmetries can be considered, i.e.,
$\gamma_0$ orbital symmetry corresponds to the choice of $n_{T\gamma_0}=1(eV)^{-1}$ and $n_{T(\gamma\ne\gamma_0)}=0$.
More realistic tips can be obtained by explicitly calculating the orbital decomposition of the tip apex PDOS in model
tip geometries, e.g., in the present work a model W(110) tip is used.

Our theory is thus an extension of the atom superposition STM/STS approach considering tunneling between directional orbitals.
The tunneling current and the differential conductance can be calculated at the tip position $\underline{R}_{TIP}(x,y,z)$ and
at bias voltage $V$ as the sum of all $\beta\leftrightarrow\gamma$ transitions between real space orbitals,
\begin{eqnarray}
I(x,y,z,V)=\sum_{\beta,\gamma}I_{\beta\gamma}(x,y,z,V),\\
\frac{dI}{dV}(x,y,z,V)=\sum_{\beta,\gamma}\frac{dI_{\beta\gamma}}{dV}(x,y,z,V),
\end{eqnarray}
respectively, with
\begin{eqnarray}
\label{Eq_i_bg}
&&I_{\beta\gamma}(x,y,z,V)=\varepsilon^2\frac{e^2}{h}\\
&\times&\sum_{a}\int_{0}^{V}T_{\beta\gamma}(E_F^S+eU,V,d_{a}(x,y,z))n_{T\gamma}(E_F^T+eU-eV)n_{S\beta}^{a}(E_F^S+eU)dU,\nonumber\\
\label{Eq_didv_bg}
&&\frac{dI_{\beta\gamma}}{dV}(x,y,z,V)=\varepsilon^2\frac{e^2}{h}\\
&\times&\left\{\sum_{a}T_{\beta\gamma}(E_F^S+eV,V,d_{a}(x,y,z))n_{T\gamma}(E_F^T)n_{S\beta}^{a}(E_F^S+eV)\right.\nonumber\\
&+&\sum_{a}\int_{0}^{V}\frac{\partial T_{\beta\gamma}}{\partial V}(E_F^S+eU,V,d_{a}(x,y,z))n_{T\gamma}(E_F^T+eU-eV)n_{S\beta}^{a}(E_F^S+eU)dU\nonumber\\
&-&\left.\sum_{a}\int_{0}^{V}T_{\beta\gamma}(E_F^S+eU,V,d_{a}(x,y,z))\frac{\partial n_{T\gamma}}{\partial U}(E_F^T+eU-eV)n_{S\beta}^{a}(E_F^S+eU)dU\right\}.\nonumber
\end{eqnarray}
This decomposition enables the analysis of the orbital contributions to the total tunneling current and to the
differential conductance. In relation to Chen's derivative rule \cite{chen90} we can state that while it is formulated inspired
by the Tersoff-Hamann model, and calculates the tunneling transmission as the absolute value square of the
tunneling matrix element that is proportional to the sample wave function derivative with respect to the real space coordinate
corresponding to the given tip orbital symmetry ($\gamma$), our transmission function also depends on the sample surface atoms'
orbital symmetry $\beta$. Moreover, the electronic structure of the tip apex is included in our theory via the PDOS.

Our method does not account for multiple scattering \cite{palotas05} and interference effects \cite{mingo96,cerda97}, which could
be important for certain systems. Therefore, it is expected that our model works well on simple metals, possibly on molecular
adsorbates on surfaces as well but not on materials with strong band structure or Fermi surface effects.
Other limitation is the uniform vacuum decay of the electron states for different orbital symmetries in
Eq.(\ref{Eq_Transmission_decomp}). This could be improved in the future.
Still, the model in its present form provides comparable results to more sophisticated tunneling models (Tersoff-Hamann, Bardeen)
as will be presented in section \ref{sec_compar}.

Note that the presented method can also be applied to magnetic systems taking into account the orbital-decomposed magnetization
PDOS of the tip and sample \cite{palotas11stm,palotas12sts} together with the orbital dependent tunneling transmission in
Eq.(\ref{Eq_Transmission_decomp}). As it was pointed out by Ferriani et al.\ \cite{ferriani10tip} the spin polarization in the
vacuum can have an opposite sign than within the tip apex atom, and this sign change is also accompanied by different dominating
orbital characters. Thus, the consideration of an orbital dependent tunneling transmission might be a better model for describing
electron transport through a magnetic tunnel junction. We return to the related spin-polarized STM/STS model in the future.

\section{Results and discussion}
\label{sec_res}

In order to demonstrate the reliability of our orbital dependent tunneling model we consider a W(110) surface. This surface has
technological importance as it is widely used as substrate for thin film growth, see e.g., Refs.\ \cite{heinze98,bode07}.
As it was pointed out by Heinze et al.\ \cite{heinze98} the determination of the position of surface atomic sites is not
straightforward as atomic resolution is lost at negative bias voltages, and a bias-dependent corrugation reversal has been
predicted. This means that normal and anticorrugated constant current STM images can be obtained in certain bias voltage ranges,
and the W atoms do not always appear as protrusions in the images. It was shown that a competition between states from different
parts of the surface Brillouin zone is responsible for this effect \cite{heinze98,heinze99}.
We reinvestigate this corrugation reversal effect as it provides a challenge for our orbital dependent tunneling model.
We find excellent agreement with the results of Ref.\ \cite{heinze98}, where an $s$-wave tip has been used, and
we study this effect in more detail.
We particularly focus on tip effects, and consider ideal tip models with different
orbital symmetries, and a more realistic W(110) tip. We find two qualitatively different corrugation inversion behaviors based
on the tip orbital composition.
Explaining our findings we highlight the role of real space orbital orientational overlaps between the surface and the tip
rather than considering electron states in the reciprocal space, thus a different kind of understanding is provided.
Finally, by comparing STM images to results of more sophisticated tunneling models we find good agreement.

\subsection{Computational details}
\label{sec_comput}

We performed geometry relaxation and electronic structure calculations based on the density functional theory (DFT)
within the generalized gradient approximation (GGA) implemented in the Vienna Ab-initio Simulation Package (VASP)
\cite{VASP2,VASP3,hafner08}. A plane wave basis set for electronic wave function expansion together with the
projector augmented wave (PAW) method \cite{kresse99} has been applied, and
the exchange-correlation functional is parametrized according to Perdew and Wang (PW91) \cite{pw91}.
The electronic structures of the sample surface and the tip have been calculated separately.

\begin{figure*}
\includegraphics[width=0.3\textwidth,angle=0]{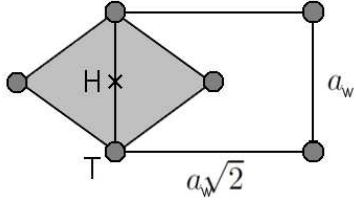}
\caption{\label{Fig2} The surface unit cell of W(110) (shaded area) and the rectangular scan area for the tunneling current
simulations. Circles denote the W atoms. The top (T) and hollow (H) positions are explicitly shown.
}
\end{figure*}

We model the W(110) surface by a slab of nine layers, where the two topmost W layers have been fully relaxed.
After relaxation the W-W interlayer distance between the two topmost layers is reduced by 3.3\%, while the underneath
W-W interlayer distance increased by 1.1\% compared to bulk W. A separating vacuum region of 18 \AA\;width in the
surface normal ($z$) direction has been set up between neighboring supercell slabs.
The average electron work function above the surface is calculated to be $\phi_S=4.8$ eV.
We used a $41\times 41\times 5$ Monkhorst-Pack (MP) \cite{monkhorst} k-point grid for obtaining the orbital-decomposed
projected electron DOS onto the surface W atom, $n_{S\beta}^{a}(E)$. The same k-set has been used for
calculating the sample electron wave functions for the Tersoff-Hamann and Bardeen simulations.
The unit cell of the W(110) surface (shaded area) and the rectangular scan area for the tunneling current simulation are shown in
Figure \ref{Fig2}. In our calculations we used the experimental lattice constant $a_W=316.52$ pm.
Moreover, the surface top (T) and hollow (H) positions are explicitly shown.

We considered different tip models. The orbital-independent ideal tip is characterized by $t_{\beta\gamma}=1$ and
$n_{T\gamma}(E)=1/9(eV)^{-1}$, so that $n_T(E)=\sum_{\gamma}n_{T\gamma}(E)=1(eV)^{-1}$.
This ideal electronically flat tip represents the limiting case of the independent orbital approximation used in previous
atom superposition tunneling models \cite{palotas11stm,palotas12sts,heinze06,palotas11sts}.
In order to study the effect of the orbital dependent tunneling other tip models are needed. 
First, we consider ideal tip models having a particular orbital symmetry $\gamma_0$.
In this case $t_{\beta\gamma}$ is calculated following Eq.(\ref{Eq_Transmission_orbdep}), and for the energy dependence of the
tip PDOS, $n_{T\gamma_0}=1(eV)^{-1}$ and $n_{T(\gamma\ne\gamma_0)}=0$ are assumed. More realistic tips can also be employed
by calculating the orbital decomposition of the tip apex PDOS in model tip geometries, and using Eq.(\ref{Eq_Transmission_orbdep})
for the orbital dependent transmission factor.
We used a blunt W(110) tip. Motivated by a previous simulation \cite{teobaldi12}, it has been modeled by a slab consisting of
three atomic layers having one W apex atom on both surfaces, i.e., with a double vacuum boundary. In this system the apex atoms
have been relaxed on both sides. The adatom-topmost layer vertical distance decreased by 19.3\% compared to bulk W.
The interaction between apex atoms in neighboring supercells in the lateral direction is minimized by choosing a $3\times 3$
surface cell, and a 17.9 \AA\;wide separating vacuum region in the $z$ direction. The local electron work function above the
tip apex was assumed to be $\phi_T=4.8$ eV. Moreover, an $11\times 15\times 5$ MP k-point grid has been chosen for
calculating the orbital-decomposed projected DOS onto the apex atom, $n_{T\gamma}(E)$.
The same k-point sampling has been used for obtaining the tip electron wave functions for the Bardeen calculation.

STM images were simulated employing our model, and the Tersoff-Hamann \cite{tersoff83,tersoff85} and Bardeen \cite{bardeen61}
methods implemented in the BSKAN code \cite{hofer03pssci,palotas05}. Scattering up to first order \cite{palotas05}
did not affect the quality of the images.
Using our model the tunneling current has been calculated in a box above the rectangular scan area shown in Figure \ref{Fig2}
containing 99000 ($30\times 22\times 150$) grid points with a $0.149$ \AA\;lateral and $0.053$ \AA\;vertical resolution.
The electron local density of states (LDOS) was calculated above the same scan area in a box of
$31\times 21\times 101$ grid points using the Tersoff-Hamann method with the same spatial resolution as above.
For the calculation of the tunneling current employing the Bardeen method a box of $31\times 10\times 100$ grid points above
the half of the rectangular scan area has been chosen in order to speed up the simulation.
In this case the lateral resolution remains $0.149$ \AA, and the vertical resolution is $0.106$ \AA.
The constant current contours are extracted following the method described in Ref.\ \cite{palotas11stm}.
All of the STM images will be presented above the full rectangular scan area.

\subsection{Convergence properties}
\label{sec_conv}

\begin{figure*}
\includegraphics[width=0.5\textwidth,angle=0]{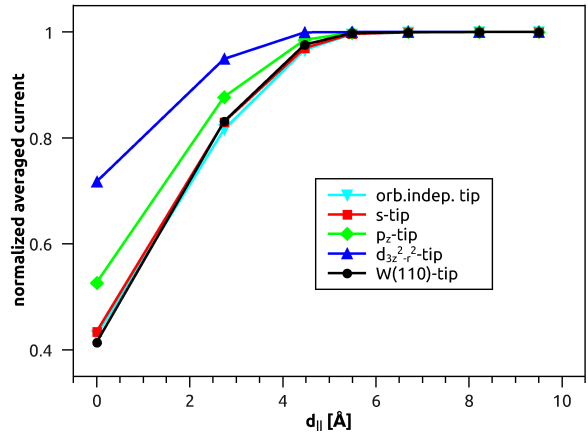}
\caption{\label{Fig3} (Color online) Convergence of the normalized averaged current calculated with different tip models.
}
\end{figure*}

Previously, the convergence of the $dI/dU$ part of the differential conductance has been investigated with respect to the
number of surface atoms involved in the summation of the orbital-independent atomic superposition formula \cite{palotas11sts}.
Due to the spherical exponential decay assumed for the electron wave functions a rapid convergence was found.
We report a similar convergence test for the orbital-dependent tunneling approach comparing different tip models.
In order to take into account a wide energy range around the Fermi level we calculated the tunneling current at
-2.5 V and +2.5 V bias voltages at $z=4.5$ \AA\;above a surface W atom, and averaged these current values.
We considered ideal tips of the orbital-independent model, and with $s$, $p_z$, and $d_{3z^2-r^2}$ symmetry, as well as
the W(110) tip. In order to obtain comparable results we normalized the averaged current for each tip calculation.
The convergences of the normalized averaged current with respect to the lateral distance on the surface, $d_{\parallel}$,
characteristic for the number of atoms involved in the atomic superposition, are shown in Figure \ref{Fig3}.
By calculating the current, contributions from surface atoms within a radius of $d_{\parallel}$ measured from the W atom
below the tip apex are summed up (sum over $a$). We find that the orbital-independent, the $s$-type, and the W(110) tips
behave quite similarly concerning the current convergence,
while for the $p_z$- and $d_{3z^2-r^2}$-type tips a faster convergence is found. This rapid convergence can be explained by the
more localized character of the latter tip orbitals in the direction normal to the sample surface ($z$). On the other hand, the
orbital-independent tip with $T=e^{-2\kappa d}$ is a good approximation for the $s$-type tip (with index $\gamma=1$), where the
spherically decaying transmission function part is still dominant, i.e.,
$T_{\beta,1}=e^{-2\kappa d}\chi_{\beta}^2$ because $\chi_{1}^2=1$.
In case of the W(110) tip, electronic states of all considered symmetries have a contribution, and their relative importance
is not only determined by the transmission function via the orbital shapes but also by the product of the symmetry-decomposed
electron PDOS of the surface and the tip. In general, the orbitals localized in different than the $z$ direction can show a slower
current convergence than the $s$ orbitals. However, the partial PDOS of such states is relatively low, and interestingly
we obtain a similar current convergence in the studied energy range as for the $s$-type tip. Choosing different bias voltages
for the W(110) tip, thus different electron states involved in the tunneling,
we found current convergences dissimilar to the $s$-type tip behavior.
The convergence can be slower or faster than obtained for the $s$-type tip depending on the partial PDOS
of each directional orbital in the given energy range.

Based on the convergence tests, atom contributions within at least $d_{\parallel}=3a\approx 9.5$ \AA\;distance from the
surface-projected tip position shall be considered.
In case of calculating STM images, $d_{\parallel}=3a\approx 9.5$ \AA\;has to be measured from the edge of the scan area in all
directions in order to avoid distortion of the image, thus involving 67 surface atoms in the atomic superposition.
For brevity, in the following we use the same surface atoms to calculate single point tunneling properties as well.

\subsection{Orbital Contributions}
\label{sec_contrib}

\begin{figure*}
\includegraphics[width=1.0\textwidth,angle=0]{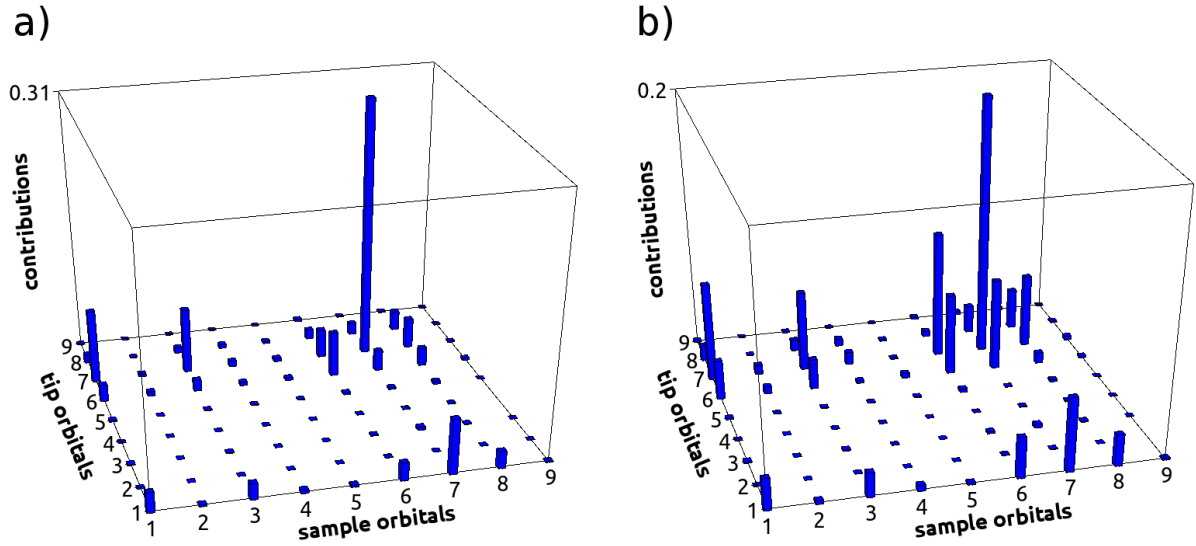}
\caption{\label{Fig4} (Color online) Histograms of the current contributions ($I_{\beta\gamma}$) from all tip-sample transitions
with different orbital symmetries using the W(110) tip at $V$= -0.1 V bias.
a) Tip apex $z=4.5$ \AA\;above the surface top (T) position (W atom);
b) tip apex $z=4.5$ \AA\;above the surface hollow (H) position, see also Figure \ref{Fig2}.
The indices of the atomic orbitals (1-9) follow the notation reported in Table \ref{Table1}.
}
\end{figure*}

Let us analyze the relative importance of all $\beta\leftrightarrow\gamma$ transitions in determining the total tunneling current
at different tip positions.
From this analysis we obtain a qualitative picture about the role of the different atomic orbitals in the construction of
the tunneling current. The $I_{\beta\gamma}$ current contributions can be calculated according to Eq.(\ref{Eq_i_bg}).
These can be represented by a current histogram that gives the percentual contributions of the individual transitions
to the total current. Figure \ref{Fig4} shows such histograms using the W(110) tip at $V$= -0.1 V bias
voltage $z=4.5$ \AA\;above two different tip positions: part a) corresponds to the tip apex above the surface top position,
and part b) to the tip apex above the surface hollow position, T and H in Figure \ref{Fig2}, respectively. We obtain a $9\times 9$
matrix from the considered orbitals, which are denoted by numbers 1 to 9 following the indices reported in Table \ref{Table1}.
We find that most contributions are due to the $s$ (1), $p_z$ (3), $d_{yz}$ (6), $d_{3z^2-r^2}$ (7), and $d_{xz}$ (8)
orbitals and their combinations. The largest contribution to the current is given by the $d_{3z^2-r^2}-d_{3z^2-r^2}$ (7-7)
transition, 31 and 20 per cent above the top and hollow positions, respectively.
Concomitantly, above the hollow position, the relative importance of both tip and sample
$d_{yz}$ (6) and $d_{xz}$ (8) orbitals is increased as it is expected from the geometrical setup, i.e., the $d_{yz}-d_{yz}$ (6-6),
$d_{yz}-d_{3z^2-r^2}$ (6-7 and 7-6), $d_{xz}-d_{xz}$ (8-8), and $d_{xz}-d_{3z^2-r^2}$ (7-8 and 8-7) contributions correspond to
larger orientational overlap of the mentioned tip and sample orbitals if the tip is above the hollow position rather than
above the top position as suggested by the geometry in Figure \ref{Fig2} and Eq.(\ref{Eq_Transmission_orbdep}).
Thus, our simple orbital dependent tunneling model captures the effect of the localized orbitals and goes beyond the spherical
Tersoff-Hamann model. Note that if a larger bias voltage is considered, i.e., the electronic states are somewhat averaged,
then the independent orbital approach might turn out to be a good approximation \cite{heinze06}.

\subsection{Atomic contrast reversal}
\label{sec_corrug}

The role of the localized orbitals can best be demonstrated by reinvestigating the corrugation inversion phenomenon found, e.g.,
on (100) \cite{mingo96}, (110) \cite{heinze98}, and (111) \cite{ondracek12} metal surfaces.
Chen explained this effect as a consequence of $m\ne 0$ tip states \cite{chen92}.
According to Heinze et al. \cite{heinze98} under certain circumstances the apparent height of W atoms at the
surface top position ($z_T$) can be larger or smaller than the apparent height of the surface hollow position ($z_H$) at
constant current ($I=const$) condition. (For the surface top (T) and hollow (H) positions, see Figure \ref{Fig2}.)
Thus, the sign change of $\Delta z(I)=z_T(I)-z_H(I)$ is indicative for the corrugation inversion.
Obviously, $\Delta z(I)>0$ corresponds to a normal STM image, where the W atoms appear as protrusions, and $\Delta z(I)<0$ to an
anticorrugated image. Since the tunneling current is monotonically decreasing with the increasing tip-sample distance, we can
obtain information about the occurrence of the corrugation inversion simply by calculating the current difference between tip
positions above the top and hollow sites of the W(110) surface. The current difference at tip-sample distance $z$ and
at bias voltage $V$ is defined as
\begin{equation}
\label{Eq_deltaI}
\Delta I(z,V)=I_T(z,V)-I_H(z,V).
\end{equation}
This quantity can be calculated for specific tips, and we call the $\Delta I(z,V)=0$ contour as the corrugation inversion map.
This gives the $(z,V)$ combinations where the corrugation inversion occurs. The sign of $\Delta I(z,V)$ provides the corrugation
character of an STM image in the given $(z,V)$ regime. Due to the monotonically decreasing character of the tunneling current,
$\Delta I(z,V)>0$ corresponds to $\Delta z(I(V))>0$, i.e., normal corrugation, and
similarly $\Delta I(z,V)<0$ corresponds to $\Delta z(I(V))<0$ and anticorrugation.

\begin{figure*}
\includegraphics[width=1.0\textwidth,angle=0]{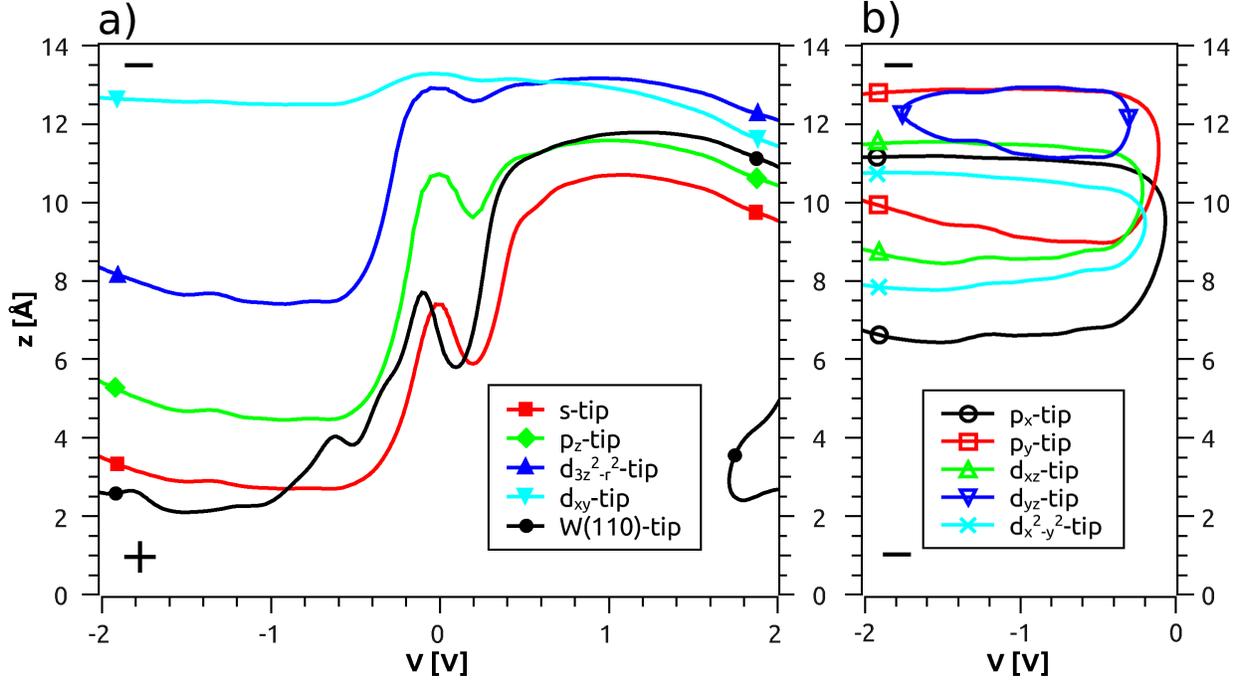}
\caption{\label{Fig5} (Color online) The $\Delta I(z,V)=I_T(z,V)-I_H(z,V)=0$ contours indicative for the corrugation inversion
[see Eq.(\ref{Eq_deltaI}), and its meaning in the text] calculated with different tip models above the W(110) surface.
Parts a) and b) show two distinct behaviors depending on the tip orbital composition.
The sign of $\Delta I(z,V)$ is explicitly shown: In part a) it is positive (+) below the curves, and negative (-) above them;
in part b) positive inside the loop of a given curve, and negative (-) outside the loop.
Note that positive $\Delta I(z,V)$ corresponds to normal, while negative to inverted atomic contrast.
}
\end{figure*}

First, we calculated $\Delta I(z,V)$ using the independent orbital approximation and Eq.(\ref{Eq_Transmission}) for the
tunneling transmission, and found that $\Delta I(z,V)$ is always positive. This means that the spherical exponential decay itself
can not account for the observed corrugation inversion effect, and the W atoms always appear as protrusions in STM images
calculated with this model. However, considering the orbital dependent tunneling transmission in Eq.(\ref{Eq_Transmission_decomp})
we find evidence for the corrugation inversion effect, thus highlighting the role of the real space shape of electron orbitals
involved in the tunneling. Figure \ref{Fig5} shows $\Delta I(z,V)=0$ contours calculated with different tip models in the
$[0$ \AA$,14$ \AA$]$ tip-sample distance and [-2 V,+2 V] bias voltage range. Before turning to the analysis of the results
obtained with previously not considered tip models let us compare our results with those of Heinze et al.\ \cite{heinze98},
where an $s$-wave tip model has been used. They found corrugation reversal at around -0.4 V at $z=4.6$ \AA\;tip-sample distance,
and above that voltage normal while below anticorrugated STM images were obtained.
Our model with an $s$-tip provides the same type of corrugation reversal at -0.21 V at the same distance as can be seen in
part a) of Figure \ref{Fig5} (curve with filled square symbol).
These bias values are in reasonable agreement particularly concerning their negative sign. At this range atomic resolution is
difficult to achieve experimentally, which is an indication for being close to the corrugation inversion regime \cite{heinze98}.
On the other hand a linear dependence of the corrugation reversal voltage and the tip-sample distance has been reported by
Heinze et al.\ $(z=4.6$ \AA$,V=-0.4$ V$)$ to $(z=7.2$ \AA$,V=0$ V$)$. Our model qualitatively reproduces this linear dependence
in the same bias range though the quantitative values are somewhat different.

Calculating the corrugation inversion maps with more tip models, we find two distinct behaviors depending on the
tip orbital composition. Parts a) and b) of Figure \ref{Fig5} show these. While the tip models in part a) can show
corrugation inversion in the whole studied bias range, this effect does not occur at positive bias voltages
for tips in part b). Moreover, anticorrugation ($\Delta I(z,V)<0$) is observed in the large tip-sample distance region
($z>13.5$ \AA) in both parts. This is in accordance with the prediction of Ref.\ \cite{heinze99} based on
the analysis of the competing electron states in the surface Brillouin zone of an Fe(001) surface.
In the $z<13.5$ \AA\;range, however, the corrugation character in the two parts of Figure \ref{Fig5} is remarkably different.
In part a), normal corrugation is found close to the surface, which reverts only once with increasing tip-sample distance for
the tip models with a single orbital symmetry in the full studied bias range. The W(110) tip behaves similarly below +1.7 V,
while above there is a double reversal of the corrugation character as the tip-sample distance increases. This indicates that
anticorrugation can be expected at short tip-sample distances (3.5 \AA-5 \AA) at around +2 V. On the other hand,
the tip models in part b) always show anticorrugation at positive bias voltages, and below -0.05 V they provide corrugation
characters starting from anticorrugation, then normal corrugation, and again anticorrugation with increasing tip-sample distance.
These different behaviors can be attributed to the tip orbital characters.
It is interesting to notice that none of the considered tip orbitals in part b) are localized in the $z$-direction, and they have
nodal planes either in the $yz$ plane ($p_x$ and $d_{xz}$) or in the $xz$ plane ($p_y$ and $d_{yz}$) or in the $x=y$ and $x=-y$
planes ($d_{x^2-y^2}$). On the other hand, in part a) there are tips which are localized in the $z$-direction
($p_z$ and $d_{3z^2-r^2}$) or having nodal planes in both the $xz$ and $yz$ planes ($d_{xy}$) as well as the spherical $s$ tip
and the W(110) tip that contains all type of orbitals with energy dependent partial PDOS functions.
The particular tip nodal planes restrict the collection of surface atom contributions to specific regions on the sample surface.
Furthermore, by changing the tip-sample distance, the orientational overlaps between the tip and sample orbitals change,
and according to our model some localized orbitals gain more importance in the tunneling contribution, see also Figure \ref{Fig4}.
Since we calculate the current difference between tip positions above the surface top and hollow sites, the complex tip-sample and
bias voltage dependent effect of the real space orbitals on the tunneling can be visualized via the corrugation inversion maps.

Concerning tips with $p_z$ and $d_{3z^2-r^2}$ orbital symmetry, Heinze et al.\ \cite{heinze98} calculated a corrugation
enhancement factor of 2 and 6.25, respectively, based on Chen's derivative rule \cite{chen90}.
Moreover, they argued that the corrugation inversion map
should be practically identical to the one obtained by using the $s$-tip model, and the corrugation values just have to be
scaled up by these factors. On the contrary based on our orbital dependent tunneling model we find that the $p_z$ and
$d_{3z^2-r^2}$ tips provide qualitatively different corrugation inversion maps, i.e., although their bias dependent shape is
similar to the one of the $s$-tip, their tip-sample distance is systematically pushed to larger values, see part a) of
Figure \ref{Fig5}. This is due to the more localized character of these tip orbitals in the $z$-direction.

Corrugation inversion with the $d_{xy}$ tip occurs at the largest tip-sample distance. A possible explanation can be based
on its $xz$ and $yz$ nodal planes. While above the top position only the underlying W atom, above the hollow position
all four nearest neighbor W atoms give zero contribution to the current, thus $I_T$ is expected to be higher than $I_H$ at
small tip-sample distances. To overcome this effect the tip has to be moved farther from the surface since then
the relative importance of the nearest neighbor contributions decays rapidly compared to other parts of the surface.

Apart from above findings we obtain corrugation inversion also in the positive bias range at enlarged tip-sample distances
for the $s$, $p_z$, $d_{3z^2-r^2}$, and W(110) tips considered in part a) of Figure \ref{Fig5}. This is most probably due to the
surface electronic structure. Note that this effect is even more difficult to capture in experiments as the corrugation values
themselves decay rapidly with increasing tip-sample distance.

\subsection{STM images - Comparison to other tunneling models}
\label{sec_compar}

In order to demonstrate the corrugation inversion more apparently, constant current STM images can be simulated. As it is clear
from Figure \ref{Fig5}, any type of crossing of the $\Delta I(z,V)=0$ contour results in the occurrence of the corrugation
reversal. In experiments two ways can be considered to record STM images in the normal and anticorrugated regimes: 1) keep the
tip-sample distance $z$ constant, and change the bias voltage $V$; or 2) keep the bias voltage $V$ constant, and change the
tip-sample distance. Respectively, these modes correspond to a horizontal and a vertical crossing of the $\Delta I(z,V)=0$ contour
in the $(z,V)$ plane in Figure \ref{Fig5}. Heinze et al.\ followed the first method in their simulations \cite{heinze98}. However,
as the second option seems to be experimentally more feasible and needs less calculations as well,
we simulated STM images at a fixed bias voltage of -0.25 V.

\begin{figure*}
\includegraphics[width=1.0\textwidth,angle=0]{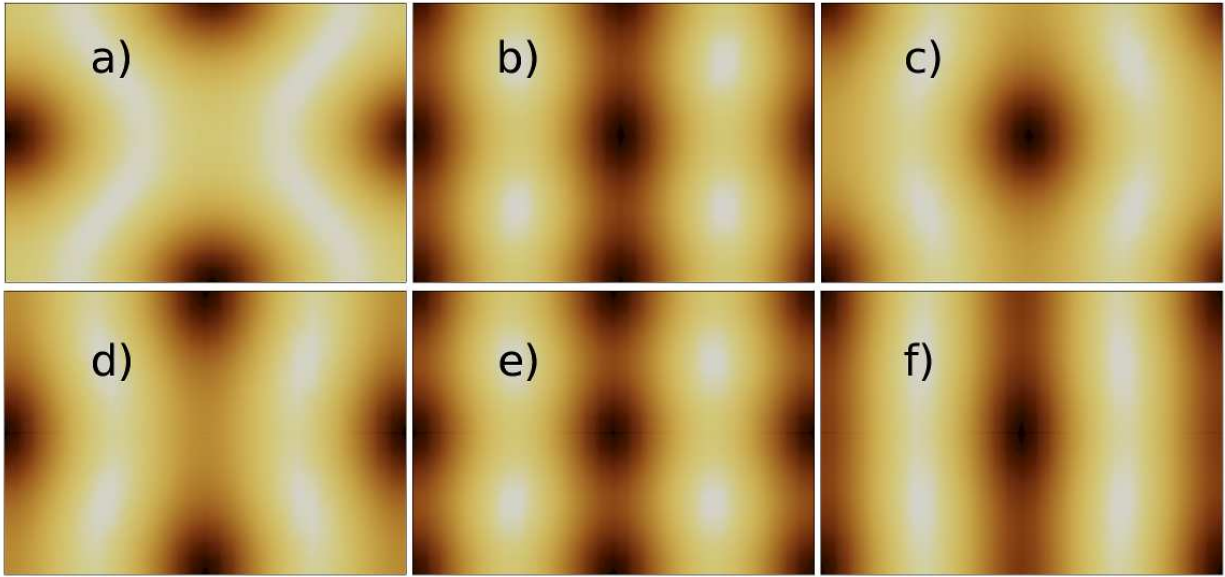}
\caption{\label{Fig6} (Color online) Comparison of simulated STM images of the W(110) surface
using our model with an $s$-type tip [top, a)-c)] and the Tersoff-Hamann model [bottom, d)-f)] at $V$= -0.25 V bias voltage.
The scan area corresponds to the rectangle shown in Figure \ref{Fig2}. Light and dark areas denote higher and lower
apparent heights, respectively. The apparent heights of the W atom ($z_T$), and the corrugation of the
contours ($\Delta z'$) are as follows:
Our model:
a) $z_T=3.80$ \AA, $\Delta z'=0.23$ pm;
b) corrugation inversion, $z_T=4.15$ \AA, $\Delta z'=0.10$ pm;
c) $z_T=4.35$ \AA, $\Delta z'=0.12$ pm.
Tersoff-Hamann model:
d) $z_T=3.80$ \AA, $\Delta z'=1.63$ pm;
e) corrugation inversion, $z_T=4.21$ \AA, $\Delta z'=1.82$ pm;
f) $z_T=4.70$ \AA, $\Delta z'=1.79$ pm.
}
\end{figure*}

In Figure \ref{Fig6} STM images are compared using our model assuming an $s$-type tip [first row a)-c)] to those calculated by the
Tersoff-Hamann method [second row d)-f)].
We find that the images are in good qualitative agreement for the a)-d), b)-e), and c)-f) pairs, respectively.
In parts a) and d), at a tip-sample distance of about 3.80 \AA, the apparent height of the W atom is larger than the one of the
hollow position, i.e.\ $\Delta z=z_T-z_H>0$. This resembles normal corrugation. Moving the tip farther from the surface,
we obtain the corrugation inversion and striped images. These are shown in parts b) and e) of Figure \ref{Fig6}.
We find that our method reproduces the corrugation inversion effect at almost the same tip-sample distance (4.15 \AA) as the
Tersoff-Hamann model predicts (4.21 \AA).
Increasing the tip-sample distance further we enter the anticorrugation regime, and the apparent height of the W atom is smaller
than the one of the hollow position, i.e., $\Delta z=z_T-z_H<0$. Such images are shown in parts c) and f).
Note that all of the simulated STM images in Figure \ref{Fig6} are in good qualitative agreement with Ref.\ \cite{heinze98}.
The corrugation of the individual current contours has also been calculated:
a) $\Delta z'=0.23$ pm, b) $\Delta z'=0.10$ pm, c) $\Delta z'=0.12$ pm,
d) $\Delta z'=1.63$ pm, e) $\Delta z'=1.82$ pm, f) $\Delta z'=1.79$ pm.
We find that our model gives approximately one order of magnitude less corrugation than the Tersoff-Hamann method.
Note, however, that the small corrugation amplitudes using our method are in good agreement with Ref.\ \cite{heinze98},
where they report $\Delta z'<1$ pm close to the contrast reversal.

\begin{figure*}
\includegraphics[width=1.0\textwidth,angle=0]{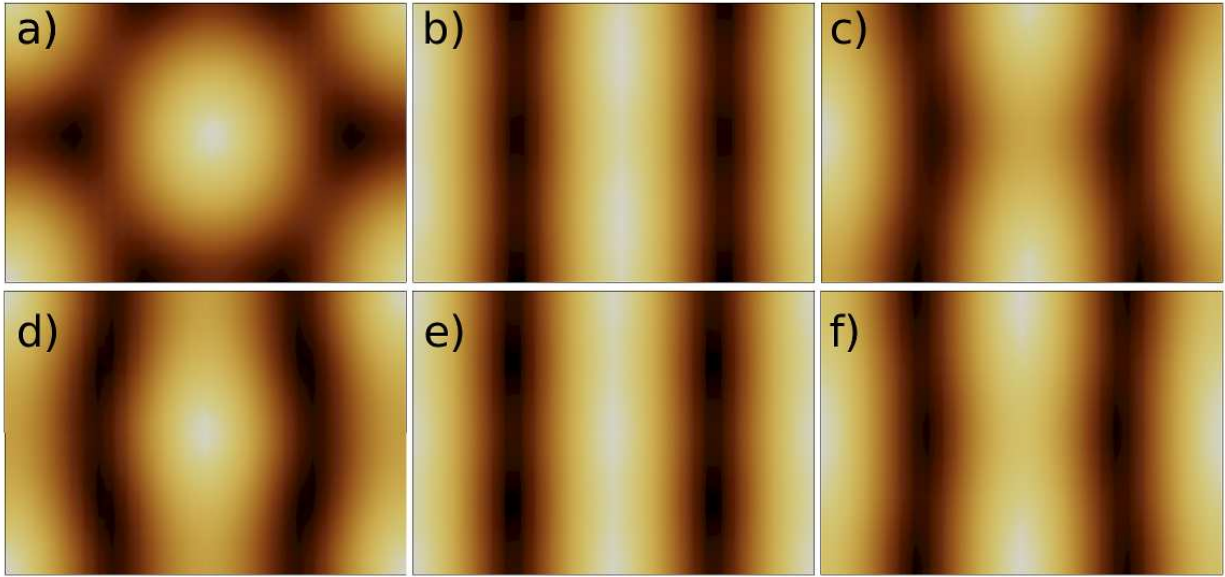}
\caption{\label{Fig7} (Color online) Comparison of simulated STM images of the W(110) surface
using our model [top, a)-c)] and the Bardeen method [bottom, d)-f)] with the W(110) tip at $V$= -0.25 V bias voltage.
The scan area corresponds to the rectangle shown in Figure \ref{Fig2}. Light and dark areas denote higher and lower
apparent heights, respectively. The current values ($I$), the apparent heights of the W atom ($z_T$), and the corrugation of the
contours ($\Delta z'$) are as follows:
Our model:
a) $I=6.3$ nA, $z_T=4.50$ \AA, $\Delta z'=0.36$ pm;
b) corrugation inversion, $I=0.43$ nA, $z_T=5.80$ \AA, $\Delta z'=0.02$ pm;
c) $I=0.35$ nA, $z_T=5.90$ \AA, $\Delta z'=0.02$ pm.
Bardeen method:
d) $I=4.4$ nA, $z_T=4.50$ \AA, $\Delta z'=0.07$ pm;
e) corrugation inversion, $I=0.7$ nA, $z_T=5.55$ \AA, $\Delta z'=0.03$ pm;
f) $I=0.19$ nA, $z_T=6.25$ \AA, $\Delta z'=0.02$ pm.
}
\end{figure*}

As we have seen, the corrugation inversion effect already occurs considering the electronic structure of the sample only.
However, Figure \ref{Fig5} indicates that different tips can modify its tip-sample distance and bias voltage dependence
quite dramatically.
In Figure \ref{Fig7} STM images are compared using our model [first row a)-c)] to those calculated by the
Bardeen method [second row d)-f)] explicitly taking the electronic structure of the W(110) tip in both cases into account.
We find that the images are in good qualitative agreement for the b)-e) and c)-f) pairs.
In parts a) and d), at a tip-sample distance of about 4.50 \AA, the agreement is weaker, however, the normal corrugation
is more pronounced in our model: The corrugation amplitude of part a) $\Delta z'=0.36$ pm is much larger than that of
part d) $\Delta z'=0.07$ pm. Moreover, as the current values of 6.3 nA (our model) and 4.4 nA (Bardeen) are comparable
to each other at the given tip-sample separation, the choice of $\varepsilon=1$ eV in Eq.(\ref{Eq_didu}) is confirmed.
Note that employing our model, a better qualitative agreement to the image of part d) has been found at a larger tip-sample
separation, i.e., closer to the corrugation inversion. This inversion is demonstrated in parts b) and e) of Figure \ref{Fig7}.
Again, we obtain striped images. Note, however, that the stripes with larger apparent height correspond to the atomic rows, in
contrast to what has been found in parts b) and e) of Figure \ref{Fig6}, where the atomic and hollow sites appeared as depressions.
This difference is definitely due to the effect of the W tip, which was not considered in Figure \ref{Fig6}. On the other hand,
we find good agreement concerning the tip-sample distance of the corrugation inversion: 5.80 \AA\;in our model, and 5.55 \AA\;
calculated by the Bardeen method. Parts c) and f) of Figure \ref{Fig7} correspond to anticorrugated images.
In this tip-sample distance regime the extremely small corrugation amplitudes are in good agreement between our model and the
Bardeen method: $\Delta z'=0.02$ pm in parts b), c), f), and $\Delta z'=0.03$ pm in part e).

Finally, we compared computation times between our model and the Bardeen method, and found the following:\\
1) our orbital dependent model, $30\times 22\times 150$ grid points, 1 CPU, time=229 s;\\
2) Bardeen method in BSKAN code, $31\times 10\times 100$ grid points, 4 CPUs, time=173866 s.\\
Normalizing to the same real space grid points we obtain that our method is 2425 times faster using 1 CPU than using 4 CPUs
for the Bardeen calculation. As the 4 CPUs calculations are roughly 3.5 times faster than the 1 CPU ones in our computer cluster,
a remarkable 1 CPU equivalent time boost of about 8500 is obtained for our method compared to the Bardeen for the given
surface-tip combination. While the k-point samplings of the surface and tip Brillouin zones affect the computation time of the
Bardeen method due to the enhanced number of transitions as the number of k-points increases, the computation time
of our model is insensitive to the number of k-points as the PDOS of the tip apex and surface atoms are used. The energy dependent
PDOS functions have the same data structure, no matter of the number of the constituting electron states obtained by the
k-summation \cite{palotas11stm}. This is a great computational advantage of our model. Of course, the quality of the results
depends on the k-point samplings. Moreover, please note the further potential that our method can be parallelized in the future.

Thus, employing our new computationally efficient orbital dependent tunneling model we could reproduce and reinvestigate
the corrugation inversion effect observed on the W(110) surface. Although this effect is driven by the surface
electronic structure, we showed that different tips can drastically modify its tip-sample distance and bias voltage dependence.

\section{Conclusions}
\label{sec_conc}

We developed an orbital dependent electron tunneling model and implemented it within the atom superposition approach
based on 3D WKB theory, for simulating STM and STS. Applying our method we analyzed the convergence and the
orbital contributions to the tunneling current above the W(110) surface. We found that the $d_{3z^2-r^2}-d_{3z^2-r^2}$
contribution is the largest, and depending on the tip position other $d$ states can gain importance as well. We also studied the
corrugation inversion effect. Using the independent orbital approximation no corrugation reversal has been obtained at all.
Employing the orbital dependent model we found corrugation reversals depending on the bias voltage in accordance with the work of
Heinze et al.\ \cite{heinze98}, and also on the tip-sample distance.
Explaining this effect we highlighted the role of the real space shape of the orbitals involved in the tunneling. Moreover,
we calculated corrugation inversion maps considering different tip models, and found two qualitatively different behaviors based
on the tip orbital composition. Our results indicate that using a W tip anticorrugation can not only be observed at negative bias
voltages but also at positive bias at reasonably short tip-sample distances. Simulation of STM images made the corrugation
inversion effect more apparent. A good agreement has been found by comparing STM images calculated by our model to Tersoff-Hamann
and Bardeen results.
The computational efficiency of our model is remarkable as the k-point samplings of the surface and tip Brillouin zones
do not affect the computation time, in contrast to the Bardeen method.
Extending this orbital dependent tunneling model to magnetic junctions is expected to enable the
study of the interplay of real space orbital and spin polarization effects in SP-STM and SP-STS experiments in the future.

\section{Acknowledgments}

The authors thank W. A. Hofer, G. Teobaldi and C. Panosetti for useful discussions.
Financial support of the Magyary Foundation, EEA and Norway Grants, the Hungarian Scientific Research Fund (OTKA PD83353, K77771),
the Bolyai Research Grant of the Hungarian Academy of Sciences, and the New Sz\'echenyi Plan of Hungary
(Project ID: T\'AMOP-4.2.2.B-10/1--2010-0009) is gratefully acknowledged. Furthermore, partial usage of the
computing facilities of the Wigner Research Centre for Physics, and the BME HPC Cluster is kindly acknowledged.

\newpage

\begin{table}
\caption{Real space orbitals, their indices used in the present paper, their definition from spherical harmonics
$Y_l^m(\vartheta,\varphi)$, and the angular dependence of their wave functions $\chi(\vartheta,\varphi)$.
Note that $\vartheta$ and $\varphi$ are the usual polar and azimuthal angles, respectively, in the spherical coordinate system
centered on the corresponding (tip or sample) atom.}
\label{Table1}
\begin{centering}
\begin{tabular}{|c|c|c|c|}
\hline
orbital & index & definition & $\chi(\vartheta,\varphi)$\tabularnewline
\hline
\hline
$s$ & 1 & $Y_{0}^{0}$ & $1$\tabularnewline
\hline
$p_{y}$ & 2 & $Y_{1}^{1}-Y_{1}^{-1}$ & $sin(\vartheta)sin(\varphi)$\tabularnewline
$p_{z}$ & 3 & $Y_{1}^{0}$ & $cos(\vartheta)$\tabularnewline
$p_{x}$ & 4 & $Y_{1}^{1}+Y_{1}^{-1}$ & $sin(\vartheta)cos(\varphi)$\tabularnewline
\hline
$d_{xy}$ & 5 & $Y_{2}^{2}-Y_{2}^{-2}$ & $sin^{2}(\vartheta)sin(2\varphi)$\tabularnewline
$d_{yz}$ & 6 & $Y_{2}^{1}-Y_{2}^{-1}$ & $sin(2\vartheta)sin(\varphi)$\tabularnewline
$d_{3z^{2}-r^{2}}$ & 7 & $Y_{2}^{0}$ & $\frac{1}{2}\left(3cos^{2}(\vartheta)-1\right)$\tabularnewline
$d_{xz}$ & 8 & $Y_{2}^{1}+Y_{2}^{-1}$ & $sin(2\vartheta)cos(\varphi)$\tabularnewline
$d_{x^{2}-y^{2}}$ & 9 & $Y_{2}^{2}+Y_{2}^{-2}$ & $sin^{2}(\vartheta)cos(2\varphi)$\tabularnewline
\hline
\end{tabular}
\par\end{centering}
\end{table}

\end{document}